\documentclass[aps,showpacs,11pt]{revtex4}
\usepackage{dcolumn}
\usepackage{graphicx}
\usepackage{amsmath}
\usepackage{amsfonts}
\usepackage{amssymb}
\usepackage{psfrag}
\usepackage{wrapfig}
\usepackage{subfigure}
\usepackage{makeidx}
\usepackage{bm}
\usepackage{epsf}
\usepackage{hyperref}
\usepackage{color}
\usepackage{multirow}

\begin{document}
	
	\title{Massive scalar field perturbation on Kerr black holes in dynamical Chern-Simons gravity}
	
	\author{Shao-Jun Zhang}
	\email{sjzhang84@hotmail.com}
	\affiliation{Institute for Theoretical Physics $\&$ Cosmology, Zhejiang University of Technology, Hangzhou 310032, China \\
	United Center for Gravitational Wave Physics, Zhejiang University of Technology, Hangzhou 310032, China}
	\date{\today}

	\begin{abstract}
		\indent We study massive scalar field perturbation on Kerr black holes in dynamical Chern-Simons gravity by performing a $(2+1)$-dimensional simulation. Object pictures of the wave dynamics in time domain are obtained. The tachyonic instability is found to always occur for any nonzero black hole spin and any scalar field mass as long as the coupling constant exceeds a critical value. The presence of the mass term suppresses or even quench the instability. The quantitative dependence of the onset of the tachyonic instability on the coupling constant, the scalar field mass and the black hole spin is given numerically.
		
	\end{abstract}
	
	%\pacs{}

	\maketitle
	
	\section{Introduction}
	
	Past years have witnessed the great achievements in astronomical observations, especially the first-ever detection of gravitational wave (GW) \cite{Abbott:2016blz,Abbott:2016nmj,Abbott:2017gyy}. With the continuous improvement of the GW detection ability, the study of black hole physics has entered a golden age, which enables us to test general relativity (GR) with unprecedented precision in the strong gravity regime. Despite the great successes of GR in explaining various astrophysical and cosmological phenomena \cite{Will:2014kxa}, there remain tension between it and quantum theory and cosmological observations. Therefore, to alleviate the tension, a variety of modified gravity theories (MOGs) have been proposed among which some were found viable by employing the current available observational constraints, such as scalar-tensor theory, dynamical Chern-Simon gravity (dCSG), scalar-Einstein-Gauss-Bonnet theory (sEGB) and Lorentz-violating gravity \cite{Berti:2018cxi}.
	
	As a special MOG, dCSG has attracted lots of attention recently. In this theory, beyond the usual Einstein-Hilbert term, an additional dynamical scalar field is introduced to couple non-minimally with the gravitational Chern-Simons invariant (also called Pontrayagin density) \cite{Jackiw:2003pm,Smith:2007jm}. This kind of coupling naturally arises in some candidate quantum gravity theories including string theory \cite{Campbell:1990fu,Moura:2006pz} and loop quantum gravity \cite{Ashtekar:1988sw,Taveras:2008yf,Mercuri:2009zi}, and also in effective field theories \cite{Weinberg:2008hq}. For a review, please refer to Ref. \cite{Alexander:2009tp}. Amounts of effort have been devoted to study black hole physics in dSCG and its astrophysical implications \cite{Cardoso:2009pk,Garfinkle:2010zx,Kimura:2018nxk,Amarilla:2010zq, Chen:2010yx, Yunes:2009hc,Konno:2009kg,Cambiaso:2010un,Yagi:2012ya,Konno:2014qua,Stein:2014xba,McNees:2015srl,Delsate:2018ome,Cunha:2018uzc,Sopuerta:2009iy,Loutrel:2018ydv,Okounkova:2019dfo}. It is interesting to note that GR black hole solutions, the Kerr black holes, are also allowed in dSCG. However, dynamics of perturbations on the same Kerr background is generally different from that in GR, which actually provides us a method to distinguish dSCG and GR through the study of perturbation dynamics. Most recently, with the presence of such coupling and in Kerr background, it is found that the massless scalar field will acquire an effective mass square which becomes negative in the vicinity of black hole horizon, resulting in the tachyonic instability and thus leading to the so-called spontaneous scalarization \cite{Gao:2018acg,Myung:2020etf,Doneva:2021dcc}. Actually, this novel phenomenon has long been observed in neutron stars but there the instability is caused by the surrounding matter instead of the curvature \cite{Damour:1993hw}. It has also been observed and studied extensively most recently in sEGB theory where a similar coupling is present but between the scalar field and the Gauss-Bonnet invariant \cite{Antoniou:2017acq,Doneva:2017bvd,Silva:2017uqg,Cunha:2019dwb,Macedo:2019sem,Doneva:2019vuh,Macedo:2020tbm,Herdeiro:2020wei,Berti:2020kgk,Dima:2020yac,Hod:2020jjy,Doneva:2020nbb,Doneva:2020kfv,Zhang:2020pko}. 
	
	The minority existing work on the tachyonic instability in dSCG  \cite{Gao:2018acg,Myung:2020etf,Doneva:2021dcc} are all focused on the case when the scalar field is massless. From the viewpoint of effective field theory, it is natural to include a mass term or a more general self-interaction term for the scalar field, which has been studied a lot in sEGB theory \cite{Macedo:2019sem,Doneva:2019vuh,Macedo:2020tbm,Doneva:2020kfv}. However, at linearized level, only the mass term can alter the onset of the tachyonic instability and the induced spontaneous scalarization. So, in the present paper, we would like to extend the study of the tachyonic instability of Kerr black hole in dCSG to the case when the scalar field is massive. We will see later that the inclusion of mass term will alter the object picture of the wave dynamics and the onset of the tachyonic instability considerably.
	
	This paper is organized as follows. In Sec. II, we give a brief introduction of the dCSG theory and write out the scalar field perturbation equation. In Sec. III, we describe our numerical method for solving the scalar field perturbation equation. In Sec. IV, we present our numerical results. The last section is devoted to summary and discussions.

	\section{Dynamical Chern-Simons gravity and massive scalar field perturbation}
	
	The action of a general dynamical Chern-Simons gravity is \cite{Alexander:2009tp,Gao:2018acg,Myung:2020etf,Doneva:2021dcc}
	\begin{eqnarray}
		S&=& \frac{1}{2\kappa}\int dx^4\sqrt{-g}\left(R-2\Lambda +\alpha f(\Phi) ^\ast R R+{\cal L}_{\Phi}\right),\nonumber\\
		{\cal L}_\Phi&=&-\frac{1}{2}\nabla^\mu\Phi\nabla_\mu\Phi-V(\Phi),\nonumber
	\end{eqnarray}
	where the scalar field $\Phi$ is non-minimally coupled to the Chern-Simons invariant 
	\begin{eqnarray}
	^\ast R R \equiv \frac{1}{2} \epsilon^{\alpha \beta \gamma \delta} R_{\alpha \beta \mu}^{~~~~\nu} R_{\gamma \delta \nu}^{~~~~\mu},
	\end{eqnarray}
	with the coupling constant $\alpha$. $f(\Phi)$ is a function of the scalar field and $\Lambda$ is the cosmological constant. From the action, one can derive the equations of motion
	\begin{eqnarray}
		\nabla^2\Phi&=&\frac{dV}{d\Phi}-\alpha  ~^\ast R R \frac{d f}{d\Phi},\label{ScalarEq}\\
		R_{\mu\nu}-\frac{1}{2}g_{\mu\nu}R + \Lambda g_{\mu\nu}&=&\alpha T^{CS}_{\mu\nu}+T^\Phi_{\mu\nu},\label{MetricEq}\\
		T^{CS}_{\mu\nu}&=& -4 \nabla^\sigma f \epsilon_{\sigma \alpha \beta (\mu} \nabla^\beta R_{\nu)}^{~\alpha} - 4 \nabla^\alpha \nabla^\beta f ~^\ast R_{\alpha (\mu\nu) \beta},\nonumber\\
		T^\Phi_{\mu\nu}&=&\frac{1}{2}\nabla_\mu\Phi\nabla_\nu\Phi-\frac{1}{2} g_{\mu\nu}V(\Phi)-\frac{1}{4} g_{\mu\nu}\nabla^\rho\Phi\nabla_\rho \Phi.\nonumber
	\end{eqnarray}
	The theory admits GR black hole solutions with constant scalar profile $\Phi=\Phi_0$,  if
	\begin{eqnarray}
		V(\Phi_0)=0,\quad \frac{dV}{d\Phi}\bigg|_{\Phi_0}=0, \quad \frac{df}{d\Phi}\bigg|_{\Phi_0}=0.
	\end{eqnarray}
	In the following, we will consider a simple case by choosing $\Lambda=0$ and $V(\Phi)=\frac{1}{2} m_\Phi^2 \Phi^2$ so that the scalar field is massive with mass $m_\Phi$ without self-interaction. Also, we assume the coupling function $f(\Phi)$ to take a general form as
	\begin{eqnarray}
	f(\Phi)=\frac{1}{2\beta} \left(1-e^{-\beta \Phi^2}\right),
	\end{eqnarray}
	where $\beta>0$ is a constant. In the small-$\Phi$ limit, $f(\Phi)$ reduces to a quadratic form considered in Ref. \cite{Gao:2018acg}.
	
	We are going to study the wave dynamics of scalar field perturbations on the background of Kerr black holes in the linear regime. The metric in the	Boyer-Lindquist coordinates is
	\begin{eqnarray}
		ds^2 = -\frac{\Delta}{\rho^2} (dt - a \sin^2 \theta d\phi)^2 + \frac{\rho^2}{\Delta} dr^2 + \rho^2 d\theta^2 + \frac{\sin^2\theta}{\rho^2} (a dt - (r^2+a^2)d\varphi)^2, \label{KerrBH}
	\end{eqnarray}
	where $\Delta\equiv r^2 - 2Mr+a^2$ and $\rho^2 \equiv r^2 + a^2 \cos^2\theta$. In this case, the scalar perturbation equation (\ref{ScalarEq}) in the Kerr background reduces to
	\begin{eqnarray} 
		&&\nabla^2\Phi=(m_\Phi^2 -\alpha ~^\ast R R) \Phi, \label{ScalarEq1}\\
		&&^\ast R R = \frac{96 a M^2 r \cos\theta (3 r^2 -a^2\cos^2\theta)(r^2-3 a^2\cos^2\theta)}{(r^2+a^2\cos^2\theta)^6},\nonumber
	\end{eqnarray}
	where the Chern-Simons (CS) invariant is valued in the background. As one can see,  the scalar field  acquires an effective mass square $m^2_{\rm eff}=m_\Phi^2 -\alpha ~^\ast R R$, which is position-dependent and may become negative close to the horizon, thus leading to possible tachyonic instability. 
	When $\alpha=0$, the above equation describes wave propagation of a free scalar field in the Kerr background which has been studied thoroughly and stability depends on the value of the physical mass $m_\Phi$: When $m_\Phi=0$, no instability is observed \cite{Krivan:1996da,PazosAvalos:2004rp,Thuestad:2017ngu}; While for $m_\Phi \neq 0$, superradiant instability may occur \cite{Damour:1976kh,Zouros:1979iw,Detweiler:1980uk,Furuhashi:2004jk,Cardoso:2005vk,Dolan:2007mj,Dolan:2012yt}.  For $m_\Phi=0$ and $\alpha>0$, tachyonic instability is found to exist for any nonzero spin as long as $\alpha$ exceeds a critical value $\alpha_c$, and $\alpha_c$ decreases as $a$ is increased \cite{Gao:2018acg}. Scalarized rotating black hole solution, which is expected to be the end-state of this tachyonic instability, has been constructed in Ref. \cite{Doneva:2021dcc} in the so-called "decoupling limit". It should be noted that the scalar field perturbation equation (\ref{ScalarEq1}) is invariant under the transformation
	\begin{eqnarray}
    \alpha \rightarrow -\alpha,\quad \theta \rightarrow \pi-\theta. \label{parity}
	\end{eqnarray}	
	So the sign of $\alpha$ can be absorbed into the CS invariant by redefining the $\theta$-coordinate, and thus the situation with $\alpha<0$ should be the same as the case with $\alpha>0$, as has been confirmed numerically and analytically in Ref. \cite{Myung:2020etf}. This is different from the case in sEGB, where $\alpha>0$ and $\alpha<0$ yield different pictures of instability and spontaneous scalarization, with the latter resulting the so-called spin-induced spontaneous scalarization \cite{Dima:2020yac,Hod:2020jjy,Doneva:2020nbb,Doneva:2020kfv,Herdeiro:2020wei,Berti:2020kgk,Zhang:2020pko}.
	
	In the following sections, taking into account the symmetry (\ref{parity}), we will only consider $\alpha>0$ and study carefully the time evolution of the massive scalar field perturbation and obtain  object pictures on the influences of the coupling constant $\alpha$ and the mass $m_\Phi$ on wave dynamics in the dCSG theory.

	\section{Numerical method}
	
	We will apply the numerical method as Refs. \cite{Krivan:1996da,PazosAvalos:2004rp,Dolan:2011dx,Doneva:2020nbb,Doneva:2020kfv} to solve the scalar perturbation equation (\ref{ScalarEq1}). In this method, the tortoise coordinate $r_\ast$ and Kerr azimuthal coordinate $\phi_\ast$ are utilized, which are defined through the transformation
	\begin{eqnarray}
	d r_\ast = \frac{r^2 + a^2}{\Delta} dr, \qquad d\phi_\ast=d\phi + \frac{a}{\Delta} dr.
	\end{eqnarray}
	In the new coordinates $(t, r_\ast, \theta, \phi_\ast)$, the semi-infinite radial domain outside the horizon $r\in (r_+, \infty)$ is mapped to infinite range $r_\ast \in (-\infty, +\infty)$ and the scalar perturbation equation can be written as
	\begin{eqnarray}\label{ScalarEq2}
	&&\left[\Delta a^2 \sin^2\theta - (r^2+a^2)^2\right] \partial^2_t \Phi + (r^2 + a^2)^2 \partial^2_{r_\ast} \Phi + 2 r \Delta \partial_{r_\ast} \Phi - 4 M a r \partial_t \partial_{\phi_\ast} \Phi\nonumber\\
	&&+ 2 a (r^2 + a^2) \partial_{r_\ast} \partial_{\phi_\ast} \Phi + \Delta \left[\frac{1}{\sin\theta} \partial_\theta (\sin\theta \partial_\theta \Phi) + \frac{1}{\sin^2\theta} \partial^2_{\phi_\ast} \Phi\right]=(m_\Phi^2 -\alpha ~^\ast R R) \Delta \rho^2 \Phi.
    \end{eqnarray}
	Taking into account the axial symmetry of the Kerr spacetime, the scalar perturbation can be decomposed as
	\begin{eqnarray}
	\Phi (t,r_\ast,\theta,\phi_\ast) = \sum_m \Psi(t,r_\ast,\theta) e^{i m \phi_\ast},
	\end{eqnarray}
	where $m$ is the well-know azimuthal number. With this ansatz and by introducing a new variable $\Pi \equiv \partial_t \Psi$, finally the perturbation equation can be cast into a form of two coupled first-order partial differential equations
	\begin{eqnarray}\label{ScalarEq3}
	\partial_t \Psi &=& \Pi,\nonumber\\
	\partial_t \Pi &=& \frac{1}{\Sigma^2}\Bigg[-4 i a m M r \Pi + (r^2+a^2)^2 \partial_{r_\ast}^2 \Psi + \left(2 i a m (r^2+a^2) + 2 r \Delta\right) \partial_{r_\ast}\Psi \nonumber\\
	& &+ \Delta \partial_{\theta}^2 \Psi + \Delta \cot\theta \partial_{\theta} \Psi  -\left(\frac{\Delta m^2}{\sin^2\theta}+m_\Phi^2 - \alpha ^\ast R R\right)\Psi \Bigg],
	\end{eqnarray} 
	where $\Sigma^2 \equiv (r^2+a^2)^2 - \Delta a^2 \sin^2\theta$. 
	
	Equations with the form as (\ref{ScalarEq3}) are suitable for the method of line \cite{Schiesser}. Precisely, the derivatives in $r_\ast$ and $\theta$ directions are approximated by finite differences, and the evolution in
	the time direction is implemented with the fourth-order Runge-Kutta integrator. Also, we impose physical boundary conditions, namely ingoing wave at the horizon and outgoing wave at infinity, following \cite{Ruoff2000}. In practical calculations, one has to truncate the infinite radial computational domain to a finite range and put boundary conditions at the outer edges, thus inevitably resulting spurious wave reflections from the outer edges. To overcome this ``outer-boundary problem", one can simply push the  outer edges to very large values so that the spurious reflections will not affect the observed signal for a sufficiently long evolution time. At the poles in 
	the angular direction $\theta=0$ and $\pi$, we impose physical boundary conditions $\Psi|_{\theta=0, \pi} = 0$ for  $m \neq 0$ while $\partial_\theta \Psi|_{\theta=0, \pi} = 0$ for  $m=0$ \cite{Dolan:2011dx}.
	 	
	As Ref. \cite{Zhang:2020pko}, the initial data of the scalar perturbation is considered to be a Gaussian distribution localized outside the horizon at $r_\ast=r_\ast^0$ and has time symmetry,
	\begin{eqnarray}
	&&\Psi (t=0,r_\ast,\theta) \sim Y_{\ell m} e^{-\frac{(r_\ast-r_\ast^0)^2}{2 \sigma^2}},\\
	&&\Pi (t=0,r_\ast,\theta)=0.
	\end{eqnarray}
	where $Y_{\ell m}$ is the $\theta$-dependent part of the spherical harmonic function and $\sigma$ is the width of the Guassian distribution. 
	In the following, without loss of generality, we take $r_\ast^0=20 M$. Also, we set $M=1$ so that all quantities are measured in units of $M$. Observers are assumed to locate at $r_\ast^0=30 M$ and $\theta=\frac{\pi}{4}$.  
	
	One should note that the Kerr spacetime is not spherically symmetric except when $a=0$, so the mode-mixing phenomenon occurs \cite{Zenginoglu:2012us,Burko:2013bra,Thuestad:2017ngu,Gao:2018acg}: a pure initial $\ell$- multipole will excite other multipoles with the same $m$ as it evolves. Taking into account this phenomenon and for simplicity, in the following we will only consider axisymmetric perturbations with $\ell=m=0$.

	\section{Results}
	
\begin{figure}[!htbp]
		\centering
		\subfigure[$~~a=0.1$ ]{\includegraphics[width=0.47\textwidth]{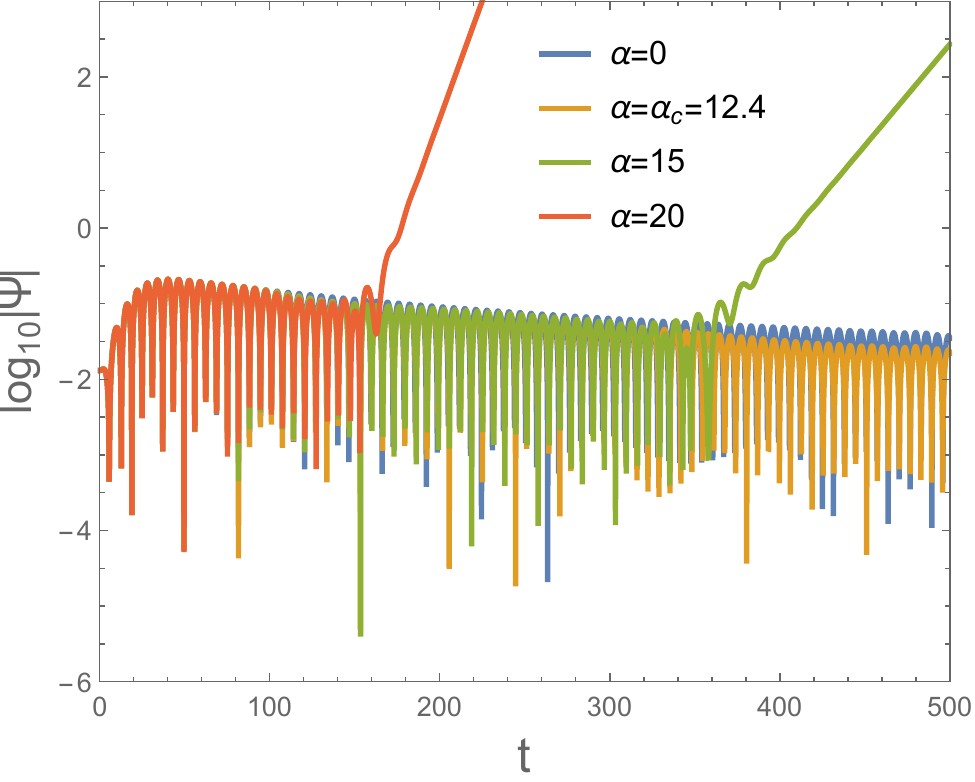}} \quad
		\subfigure[$~~a=0.8$ ]{\includegraphics[width=0.47\textwidth]{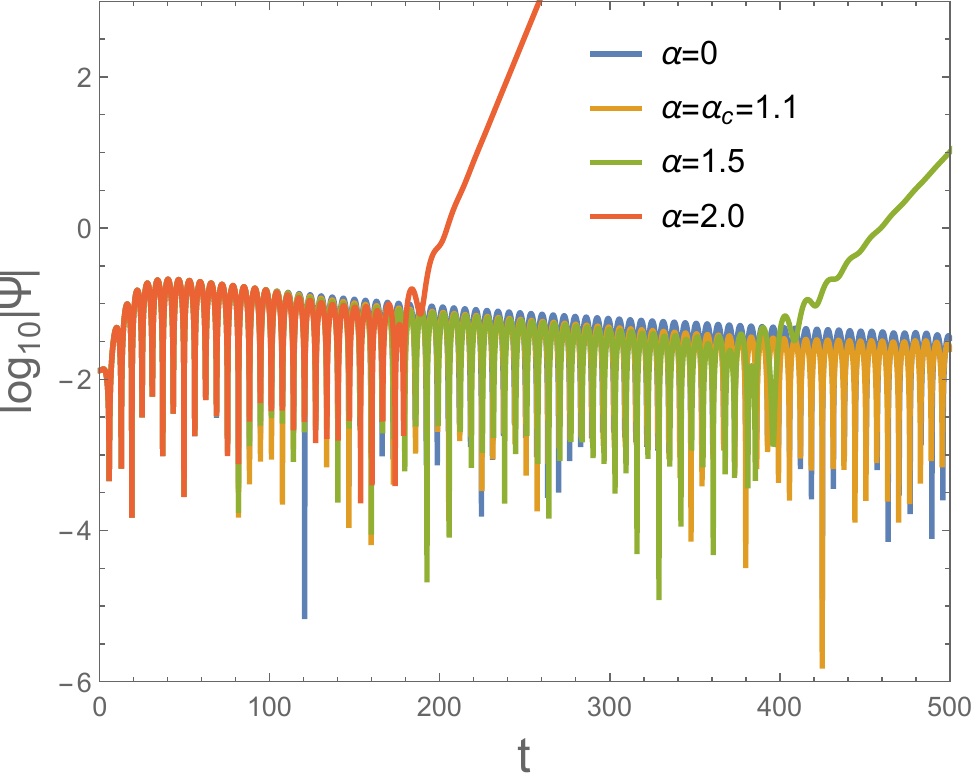}}
		\caption{(color online) Time evolution of the scalar perturbation for $a=0.1$ and $0.8$ with fixed scalar field mass $m_\Phi=0.5$. The initial multipole we considered is $\ell=m=0$. Time is in units of $M$.}
		\label{PsiM05}
	    \end{figure}

\begin{figure}[!htbp]
    	\centering
    	\subfigure[$~~a=0.1, \alpha=30$ ]{\includegraphics[width=0.47\textwidth]{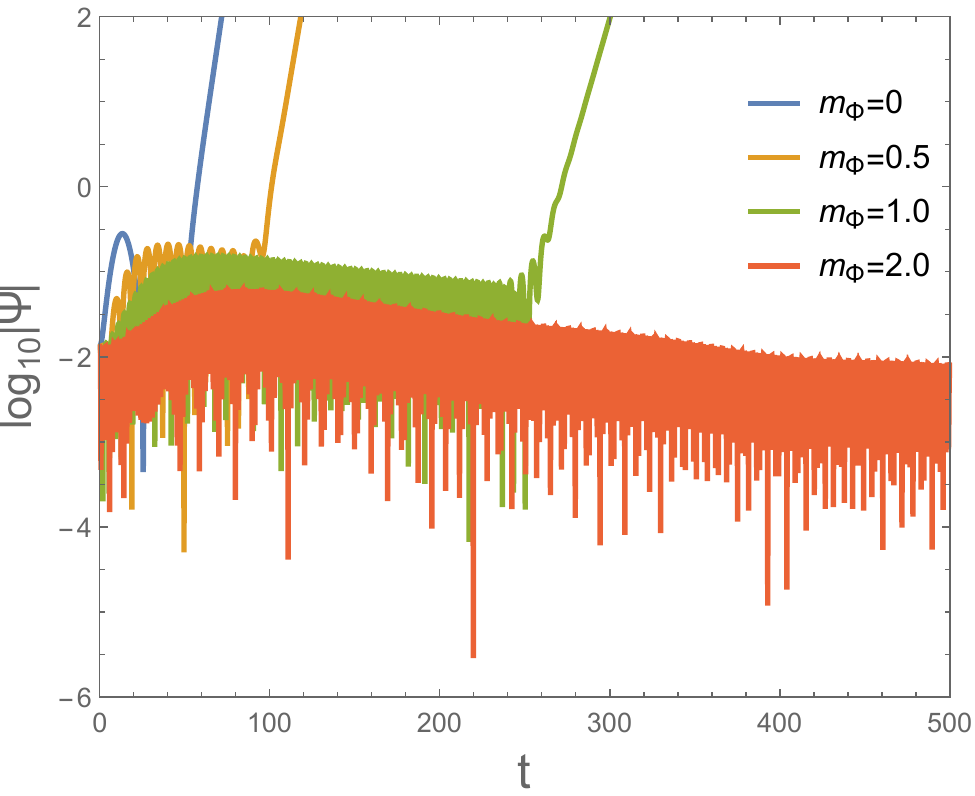}} \quad
    	\subfigure[$~~a=0.8,\alpha=3.0$ ]{\includegraphics[width=0.47\textwidth]{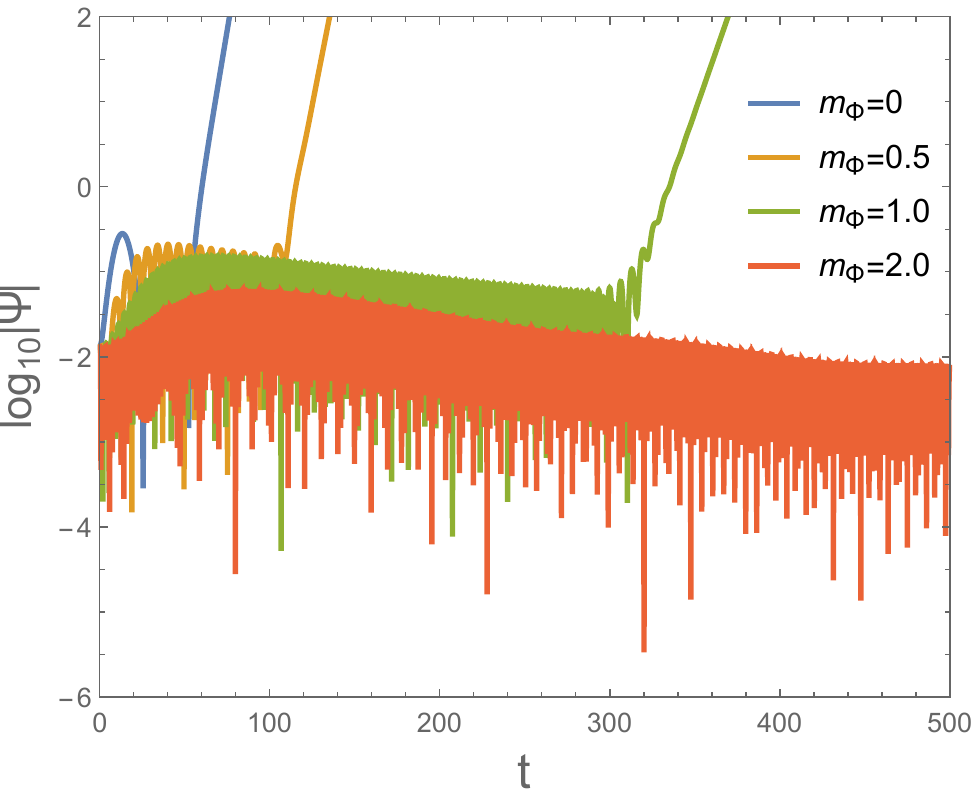}}
    	\caption{(color online) Time evolution of the scalar perturbation. The parameters are fixed as $(a=0.1, \alpha=30) $ ({\em left}) and $(a=0.8, \alpha=3.0)$ ({\em right}). The initial multipole we considered is $\ell=m=0$. Time is in units of $M$.}
    	\label{Psia0108}
    \end{figure}

	We have performed the time evolution of the scalar perturbations for various values of spin $a$ and scalar field mass $m_\Phi$, and found that instability always occurs as long as the coupling constant $\alpha$ exceeds a critical value $\alpha_c$. Representative examples are given In Figs. \ref{PsiM05} and \ref{Psia0108}, with similar pictures for other values of parameters.
		
   In Fig. \ref{PsiM05}, time evolutions of the axisymmetric scalar perturbation are plotted with fixed spin and scalar field mass. From the figure, one can see that instability will be triggered once the coupling constant $\alpha$ exceeds a critical value $\alpha_c$, and $\alpha_c$ decreases as the spin $a$ is increased. For $\alpha>\alpha_c$, larger $\alpha$ makes the instability to appear earlier and more violent. In Fig. \ref{Psia0108}, we fix the spin and the coupling constant to study the influence of the scalar field mass on the time evolution of the perturbations. From the figure, one can see that increasing $m_\Phi$ will suppress the instability and delay its appearance, and the instability will cease to appear if $m_\Phi$ is further increased, which implies that $\alpha_c$ increases as $m_\Phi$ is increased.

\begin{figure}[!htbp]
   	\centering
   	\subfigure[$~~a=0.1$ ]{\includegraphics[width=0.47\textwidth]{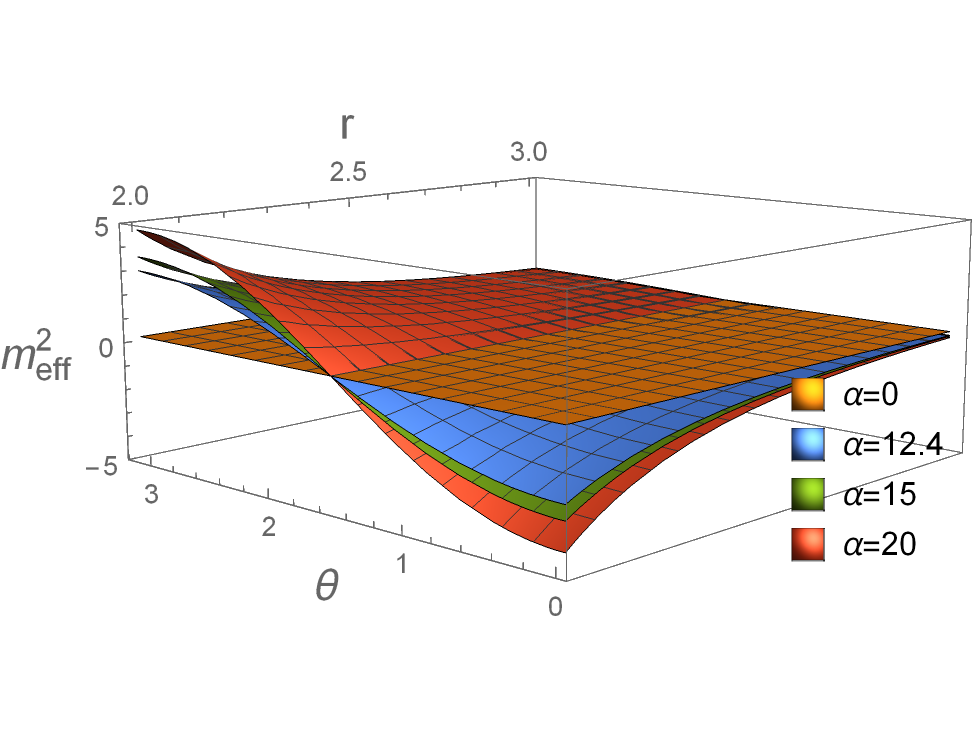}} \quad
   	\subfigure[$~~a=0.8$ ]{\includegraphics[width=0.47\textwidth]{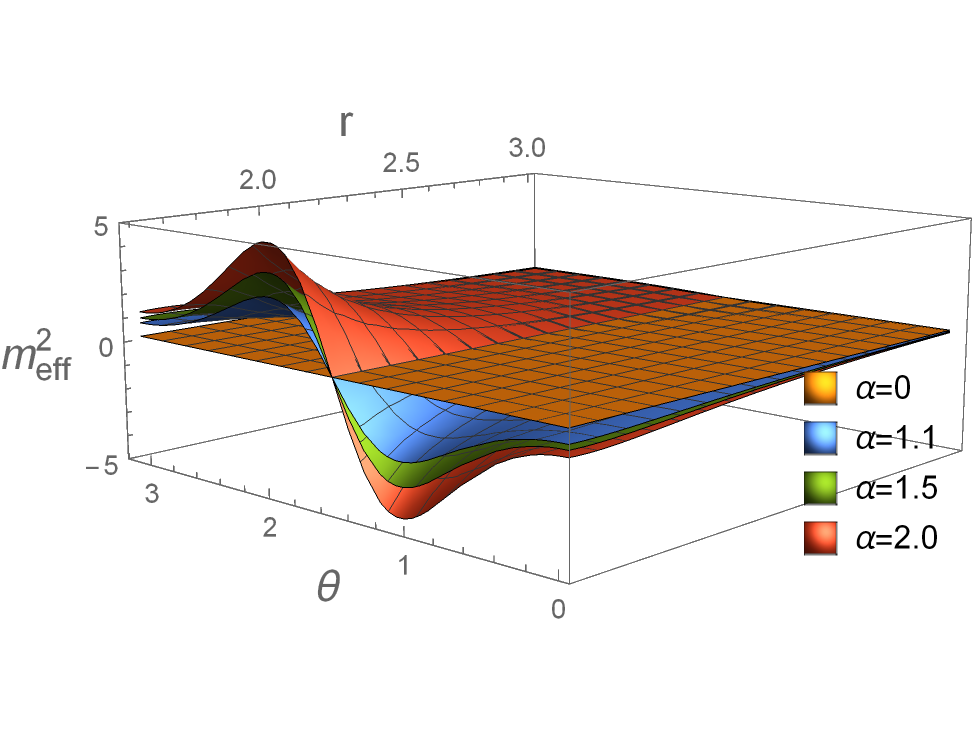}}
   	\caption{(color online) Profiles of the effective mass square $m^2_{\rm eff}$ for $a=0.1$ and $0.8$ with fixed scalar field mass $m_\Phi=0.5$.}
   	\label{EffectiveMassM05}
   \end{figure}

   Physically, the influences of the coupling constant and the scalar field mass on the stability can be understood qualitatively from the profiles of the effective mass square $m^2_{\rm eff}=m_\Phi^2 -\alpha ~^\ast R R$, as shown in Fig. \ref{EffectiveMassM05}. The profiles exhibit odd parity under the transformation $\theta \rightarrow \pi -\theta$ as we have already mentioned above in Eq. (\ref{parity}). We should note that the effective mass square is a positive constant $m^2_{\rm eff}=0.5^2$ when $\alpha=0$. As $\alpha$ is increased and exceeds some value $\alpha_0$, $m^2_{\rm eff}$ will become negative in vicinity of the horizon for $\theta \in [0,\frac{\pi}{2})$ and become more negative when $\alpha$ is further increased. We note that $\alpha_0<\alpha_c$, which implies that small negative $m^2_{\rm eff}<0$ is not sufficient to trigger tachyonic instability. Only when $m^2_{\rm eff}$ is negative enough ($\alpha>\alpha_c$) can the instability be developed. With the further increase of $\alpha$, the instability appears earlier and becomes more violent for which $m^2_{\rm eff}$ will become more negative. Moreover, from the analytical definition of the effective mass square, it is explicit that the influence of the scalar field mass $m_\Phi$  is opposite to that of the coupling constant.
   
\begin{figure}[!htbp]
		\centering
		\includegraphics[width=0.47\textwidth]{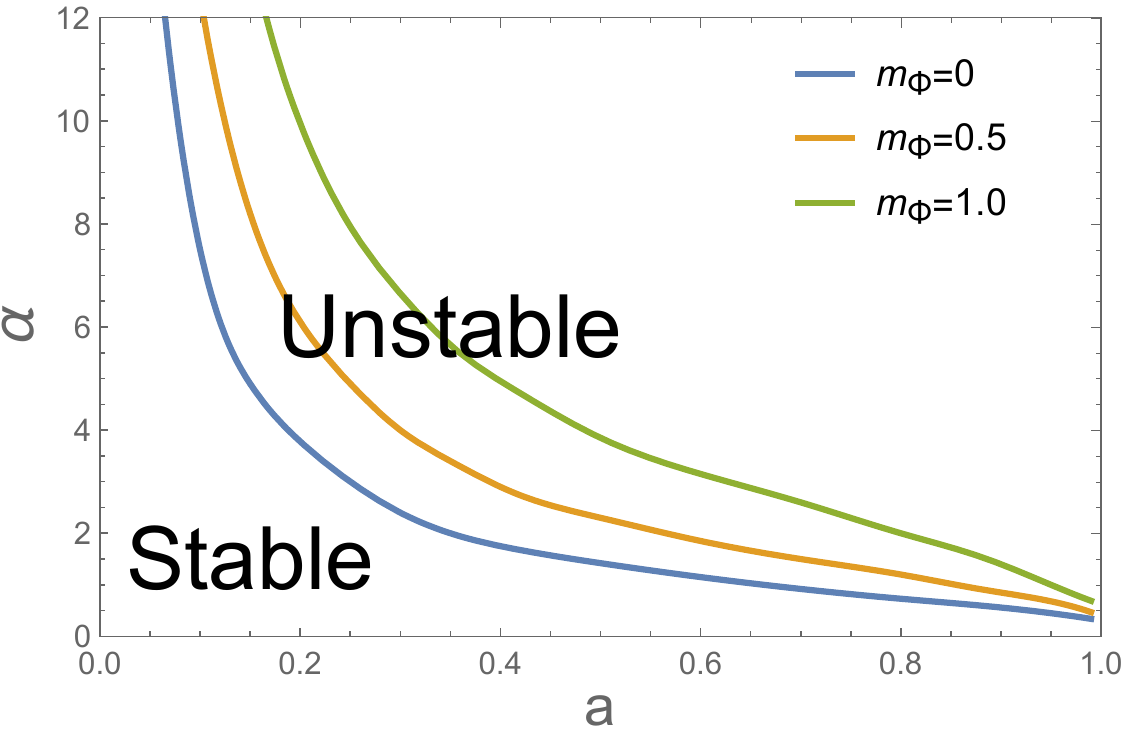}
		\caption{(color online) Boundary between stable and unstable regions in $a-\alpha$ plane for different scalar field masses. The initial multipole we considered is $\ell=m=0$. }
		\label{UnstableRegion}
	\end{figure}

   The more complete picture of the influences of the parameters $(a, \alpha, m_\Phi)$ on the onset of the tachyonic instability is summarized in Fig. \ref{UnstableRegion}, from which the above mentioned phenomena can be seen more clearly. When $m_\Phi=0$, the scalar field becomes massless and its wave dynamics has been studied in Ref. \cite{Gao:2018acg} by adopting a different numerical strategy. Our results for this particular case are in good agreement with those there.

	\section{Summary and Discussions}
	
	In this work, within the framework of dCSG theory, we studied carefully the time evolution of the massive scalar field perturbation on Kerr background by performing a $(2+1)$-dimensional simulation. We found that tachyonic instability always occurs for any nonzero spin $a$ and any scalar field mass $m_\Phi$ as long as the coupling constant $\alpha$ exceeds a critical value $\alpha_c$. The value of $\alpha_c$ depends on the values of $a$ and $m_\Phi$. For fixed $m_\Phi$, $\alpha_c$ decreases as $a$ is increased; While for fixed $a$, $\alpha_c$ increases as $m_\Phi$ is increased, which means the scalar field mass $m_\Phi$ will suppress the instability or even quench the instability if $m_\Phi$ is large enough. Physically, as shown in Fig. \ref{EffectiveMassM05}, the influences of the parameters $\alpha$ and $m_\Phi$ on the onset of the tachyonic stability can be explained qualitatively from the behaviors of the effective mass square $m^2_{\rm eff}=m_\Phi^2 -\alpha ~^\ast R R$.
	
	Although, we have obtained object pictures of the time evolution of the scalar field perturbation and the quantitative influences of the parameters $(a, \alpha, m_\Phi)$ on the onset of the tachyonic instability, there remains several interesting issues. From Figs. \ref{PsiM05} and \ref{Psia0108}, one can see that, if instability is not triggered $(\alpha<\alpha_c)$, the scalar field perturbation will exhibit oscillatory behavior at late time. Similar behavior has already been observed in GR $(\alpha=0)$ \cite{Price:1971fb,Burko:2004jn} and sEGB \cite{Doneva:2020kfv} theories, and also in dCSG theory for spherically symmetric black hole background \cite{Macedo:2018txb}, with analytical expression as	
	\begin{eqnarray}
	\Psi \sim \cos (\omega_c t) t^{p},
	\end{eqnarray}
	where $\omega_c \sim m_\Phi$. The power-law index $p$ exhibits transitional behavior from $p=-(\ell +3/2)$ at intermediate times to $p=-5/6$ at very late times. From  the figures, it is interesting to see that the coupling constant $\alpha$ nearly has no influence on such oscillatory behavior.  How to understand this phenomenon needs more careful studies on the late time tail. Beyond the scalar field perturbation we considered in present work, there is of course the possibility of gravitational field perturbation as well. The coupling between the two types of perturbations may make the phenomena richer. This issue has been studied extensively for spherically symmetric black hole background in Refs. \cite{Molina:2010fb,Pani:2011xj,Macedo:2018txb}. It is interesting to extend these studies to rotating case to see the influence of the black hole spin. In this work, we only study the time evolution of the scalar field perturbation in linearized level. The appearance of instability indicates the possible existence of scalarized black holes as an end-state. To gain better understanding of the fate of this instability, a full non-linear evolution of the perturbation and the construction of scalarized black holes are called for. As perturbations with higher azimuthal number $m$  normally trigger more moderate instability, so in this work we only focus on axisymmetric perturbations with $m=0$. For perturbations with $m \neq 0$, beyond the tachyonic instability, there may appear another type of instability for massive scalar field perturbations, the well-known superradiant instability \cite{Brito:2015oca}. Unfortunately, as the growing time of the superradiant instability is usually very large, the observation of this instability requires a long stable time evolution of the perturbations, which will be a great challenge for numerical calculations. We leave these questions for further investigations.

	\begin{acknowledgments}
		
		This work is supported by the National Natural Science Foundation of China (NNSFC) under Grant No 12075207.
	\end{acknowledgments}

\end{document}